# DCarbonX Decentralized Application: Carbon Market Case Study

Nida Khan [*] and Tabrez Ahmad [†]

[*]Nash fintechX, Luxembourg
Email: nida.khan@nashfintechx.com

[†]ArcelorMittal, Luxembourg
Email: tabrez.ahmad@arcelormittal.com

**Abstract**

Decentralized applications developed using blockchain technology provide innovative business models to serve the human race and solve existing challenges. Climate change is one of the biggest problems humanity is facing and there is a dearth of solutions in tackling this grave impediment to the long-term sustainability of our planet. Accountability, greenwashing, traceability, impact assessment and trading of carbon credits are unresolved issues in the ESG sector. In this paper, we present a novel decentralized application software, DCarbonX, that solves the enumerated problems using NFTs on the blockchain platform, through smart contracts. The paper describes the functional architecture of DCarbonX, while elaborating on its salient features and utility in sustainable finance, in particular green sukuk. DCarbonX is a pioneering software providing an exchange for trading of carbon credits. The software facilitates logging of impact and traceable transactions in a carbon market, that would help to prevent duplication of records and greenwashing. The paper discusses the efforts being undertaken to achieve the climate goals as per the Paris Agreement and also highlights the pivotal obstacles to achieving carbon neutrality by 2050, as per COP26. The paper also encompasses a study on the applications of dapps in DeFi, Web 3.0 and ESG, among other areas and gives a comparative analysis of blockchain platforms for dapp development. The paper is also a pioneer in highlighting the challenges that plague dapp development, deployment and usage.

## I. INTRODUCTION

The advent of blockchains in 2008 [1] with a novel concept of economic incentives for participation in network reliability initiated the domain of cryptoeconomics. Blockchain technology garnered considerable interest from domains spanning from finance [2] to education [3], with a potential to disrupt existing working models. The pandemic caused by the novel coronavirus has established the need for a strong digital infrastructure for governments to ensure that the effects on the economy are minimal [4]. Blockchain can assist in deploying this digital infrastructure while providing transparency and data integrity. The financial sector is seen as the pivot for blockchain to revolutionize the economy by innovative models of payments [5] and banking [6]. Blockchain evolved from a permissionless network in Bitcoin to a permissioned one in Quorum, resonating with the existing legacy financial systems to be in synchronization with the needs for data privacy and regulated access. The conception of smart contracts in Ethereum [7] heralded a new era of applications, built on top of the underlying decentralised blockchain infrastructure, known as decentralised applications (hereafter referred to as dapps).

Dapps utilize and provide the characteristics of blockchain technology to products and services, majorly using smart contracts [8]. Smart contracts find applicability in multiple areas, including but not limited to Islamic capital markets [9], peer-to-peer lending [10], supply chain finance [11], insurance [12] and the Environmental, Social and Governance (ESG) sector [13]. The market for dapps is expected to reach 368.25B by 2027 spurred by the need for secure peer-to-peer transactions, where coronavirus is seen as a key influencing factor [14]. The primary advantages provided by dapps include transparency, immutability, real time transactions, low cost, absolving of intermediaries and a programmed execution of stipulated terms in contractual business relationships. Dapps through smart contracts can also help to adhere to Shariah standards by helping to comply with the Shariah law and providing evidence for the same through transparent transactions on the blockchain [15].

The ESG market is facing multiple challenges like lack of transparency, issues dealing with integration in the existing infrastructure, and delivering on ESG commitments [16]. A pertinent area of relevance after COP26 is that of carbon emission reductions and countries are being asked to come forward with ambitious 2030 reductions targets to reach *"global net zero by mid-century and keep 1.5° within reach"* [17]. A major requirement to achieve this goal is to have an authentic and accurate record of carbon emissions by each country and hence each relevant organisation. Thereafter, there should be an infrastructure to provide this record to the country for accounting purposes such that the data is non-repudiated and cannot be corrupted. Moreover, an exchange of carbon credits should be facilitated to ensure cooperation between organisations encouraging flow from carbon negative entities to carbon neutral and carbon positive entities. The paper conducts a case study on DCarbonX, which is a pioneering blockchain-based software, that solves some of the major problems associated with tracking and tracing of carbon credits. DCarbonX is developed by Nash fintechX [18], which is a software solutions provider specialising in blockchain and artificial intelligence, to provide an accurate recording of carbon credits and a trading market for exchange of carbon credits. The case study has been discussed in Section IV. The case study elaborates on the motivation behind the development of DCarbonX, the functional architecture, salient features and the utility of the developed decentralized application software in sustainable finance.

The rest of the paper is structured as follows. The related work is given in Section II. Relevant background, where a description of blockchain technology, smart contracts, dapps and a comparison of major blockchain platforms on which dapps can be developed is given in Section III. The paper discusses the applications of dapps in some pertinent domains in Subsection III-E. The challenges faced during the development, deployment and usage of dapps, in general, are given in Section V while the conclusion is given in Section VI.

## II. RELATED WORK

Cai et al. in [19] elaborate on the importance of dapps and conduct a survey on state of the art in dapps, whereas in [20], Zheng et al. discuss the components of a dapp, while highlighting that there is no single point of failure. In this paper, we elaborate on the issues that are present in the ESG domain, while providing a solution to solve some major impediments to achieving carbon neutrality by 2050. In [21], Wu et al. do an empirical study of blockchain-based dapps, provide a summary of methodology of usage of smart contracts by dapps to access the underlying blockchain and conduct a descriptive analysis of popularity of dapps. The present work differs in focusing on a single decentralized application software, DcarbonX, that provides a carbon market for trading of carbon credits. In [22] Mishra et al. discuss the usage of blockchain and dapps to resolve the security issues around sharing of students' credentials, whereas our work deals with resolving issues around transparency, tracking and accurate accounting of carbon emission reductions. Rupa et al. in [22] utilize a dapp for storage of medical records in Ethereum to protect it from unauthorized access by attackers and designed the associated dapp for Industry 5.0. The present work utilizes a dapp to help prevent greenwashing and manipulation of the record of issued carbon credits. The present work is an application for Web 3.0. Marchesi et al. in [24] define a structured approach to developing dapps using an agile and scrum methodology in Ethereum, which is extensible to other blockchains. Our work defines the architecture and gives the platform details for DCarbonX, which is a dapp providing a blockchain-based carbon market. In [25], Hamza et al. give an overview of the scientific evidence and impact of climate change, indicating the need for solutions to combat it. The present work is a pioneer in providing a solution through a dapp, DCarbonX, for assisting in providing an accurate quantitative measurement of the efforts to combat climate change. In [26], Chen elaborates on the significance of using blockchain to improve accountability in carbon markets and to develop renewable energy microgrids, whereas our work provides a practical evidence of the accountability blockchain provides for carbon markets through our decentralized application software, DCarbonX.

## III. BACKGROUND

*A. Blockchain*

Blockchain is a decentralized and distributed database that operates in a peer-to-peer network. The organisation of data in a blockchain is in the form of blocks and can be visualized as a singly linked list as given in Figure 1. The singly linked list is a series of nodes, where each node contains a data field and the address of the next data node in the list. The last node in the list points to null. Blockchain is an ever increasing series of blocks, where the first block is referred to as the genesis block and provides a template for all other blocks. All blocks, other than the genesis, contain a reference to the previous block. There are conceptual differences between a singly linked list and blockchain. Individual blocks in the blockchain have multiple data points in the form of transactions hashed together in a Merkle tree and hash of the previous block in the block header, for example, linked together by cryptography and this is not present in a singly linked list. Additionally, the data in a linked list can be changed and deleted. In a blockchain, the data is immutable and cannot be corrupted. There is no provision for deleting blocks in a blockchain as any change in data needs to be accomplished in all the copies of the database that reside with possibly thousands of validators in case of a public blockchain [1]. The blockchain network majorly comprises of:

- Users
- Validators
- Consensus Protocol
- Governance Mechanism

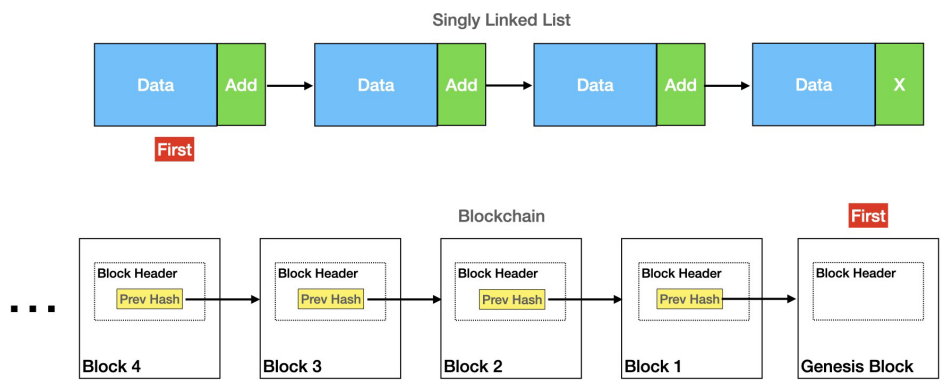

*Figure 1: Singly linked list vs Blockchain.*

Users use the dapps and payment services provided by the blockchain. Validators are responsible for maintaining the blockchain database and participate in validating transactions by virtue of which the data gets recorded on the distributed ledger of the blockchain. Consensus protocol has been described in the following subsection. The governance mechanism followed by the blockchain network determines how coordinated the various entities in the network are to accomplish the strategic goals of the blockchain platform, ensures that the blockchain services are always available and minimises the occurrence of forks [27].

*1) Consensus:* Blockchain is a distributed database where no single entity controls the addition of data but the right to update the database is distributed among owners of computing power, stakeholders or a user's social network [28]. The universal agreement within the blockchain network as to the legitimacy of the data to be added and which validator will add data is referred to as consensus. There are different consensus protocols being utilized by blockchain networks based on the economic set and the choice of the consensus protocol helps in determining the throughput of the blockchain network, among other major parameters governing blockchain networks. The primary consensus protocol, *Proof of Work*, employed by Bitcoin and Ethereum, utilizes a lot of energy and the energy consumption of each Bitcoin transaction is estimated to be equivalent to 1173 kilowatt hours of electricity, which can *"power the typical American home for six weeks"* [29]. This makes proof of work consensus protocol an unsuitable candidate for usage in view of the climate goals outlined in COP26.

*B. Smart contracts*

Smart contracts are computer programs that reside on the blockchain platform and their correct execution is enforced by the consensus protocol. They help to formalize and secure relationships on the blockchain network [30]. Smart contracts find usage in multiple areas like gaming, finance, supply chain, insurance and notary, among others [31]. A sample smart contract to create an NFT (non-fungible token), which are token representations of non-fungible assets, is given in Figure 2. The smart contract creates a permanent record of an Eid Al Adha 2021 greeting by Nash fintechX, which depicts a few lessons from the last sermon of Prophet Mohammad (Sallalahu Alayhi Wasallam - May peace and blessings of Allah be upon him). The smart contract stores the hash of the greeting image that is stored on InterPlanetary File System (IPFS), which is a peer-to-peer distributed file system [32]. Nash fintechX made use of Pinata cloud services [33] to ensure the image can be accessed at all times from IPFS. Pinata provides pinning services to facilitate a user to host files on the IPFS network. The smart contract is stored on the Rinkeby testnet of Ethereum.

The transaction hash for the NFT from Nash fintechX is:

*0x136d93be4784b3b4c470337608a8aa1b5f2402fa3edb15128a57a6902669e108*

The transaction hash can be searched in the Rinkeby testnet explorer [34]. Thereafter, the transaction details can be seen to verify input data. The input data can be decoded to see the link for the file stored on IPFS, which in turn contains the link of the image and other relevant information associated with the NFT. The image that is stored on IPFS is given in Figure 3.

```solidity
1   pragma solidity 0.8.6;
2
3   import "https://github.com/0xcert/ethereum-erc721/src/contracts/tokens/nf-token-metadata.sol";
4   import "https://github.com/0xcert/ethereum-erc721/src/contracts/ownership/ownable.sol";
5
6   contract newNFT is NFTokenMetadata, Ownable {
7     // declare the total supply of the NFT
8     uint256 public _totalSupply = 1;
9     // function to initialize the name of the NFT and the symbol
10    constructor() {
11      nftName = "NashfintechX NFT";
12      nftSymbol = "NashX";
13    }
14    // function to mint the NFT passing the receiver's address, token ID and the url of the NFT json file
15    function mint(address _to, uint256 _tokenId, string calldata _uri) external onlyOwner {
16      super._mint(_to, _tokenId);
17      super._setTokenUri(_tokenId, _uri);
18    }
19  }
```

*Figure 2: An Excerpt from an NFT Smart Contract*

## C. Decentralized applications

Decentralized applications or dapps, as mentioned previously, are developed on the blockchain infrastructure supported by smart contracts. The nomenclature is derived from the way existing applications are designated using the environment of their execution as a deciding factor. Hence, applications developed to function on the web are known as web apps whereas those that function on smartphones are referred to as smartphone apps. Similarly applications that function on decentralized platforms like the blockchain are referred to as decentralized apps or dapps. The difference is that dapps have a backend hosted on a decentralized platform, whereas web apps and smartphone apps are powered by a centralized backend in AWS for example. Ethereum was the first blockchain platform that provided the functionality of smart contracts [35] in 2015. Majority of dapps are on Ethereum [36] with the number being close to 4000 [37].

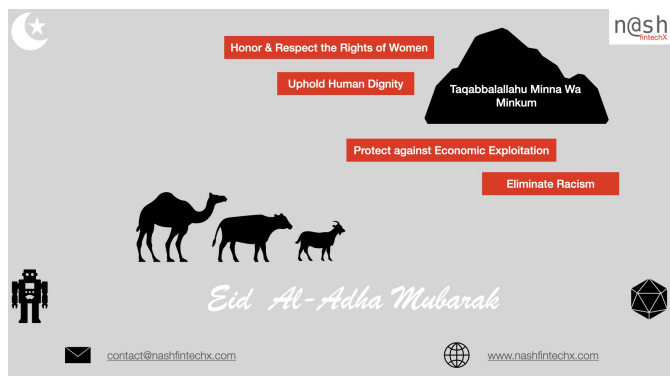

*Figure 3: Eid Al Adha Greeting 2021 from Nash fintechX - NFT stored on Rinkeby testnet of Ethereum.*

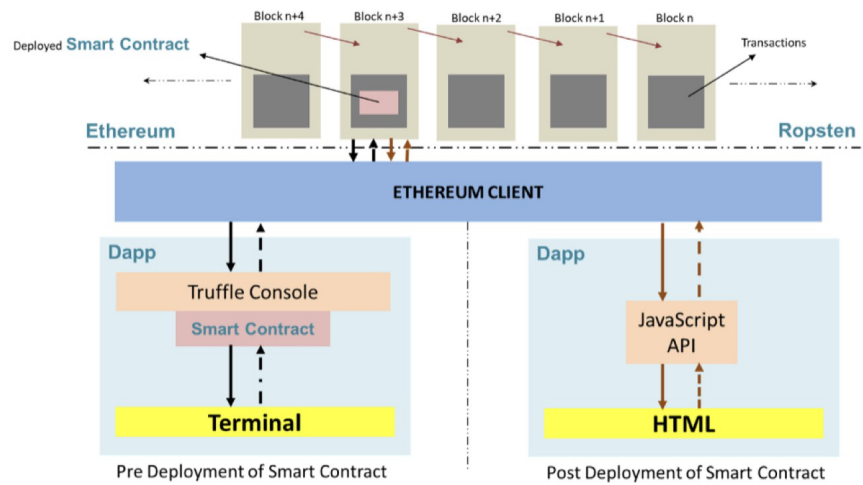

*Figure 4: Ethereum Dapp Development.*

Figure 4 depicts one of the methodologies of dapp development in Ethereum. The smart contract/s on which the dapp would be developed are deployed on Ethereum using Truffle Console [38] through the Linux command line [39]. Truffle console helps to deploy the smart contract on the blockchain through an Ethereum client [40]. Once the smart contract has been deployed, the user interface (UI) is developed to help the users interact with the smart contract on the blockchain. In Figure 4, the smart contract has been deployed on the Ropsten testnet of Ethereum, which simulates the main Ethereum network but offers fake cryptocurrencies for testing purposes. The frontend helps the user to interact with the smart contract through the Ethereum client using Ethereum JavaScript API [41]. All the changes in data through the dapp are recorded using the smart contract on the blockchain platform. The methodology followed in different blockchain platforms by way of software libraries and tools used might be different from the one depicted in Figure 4, but conceptually the process will be similar in all blockchain platforms.

*D. Blockchain Platforms for Dapp Development*

Blockchain platforms that enable the functionality of smart contracts can provide the infrastructure to develop dapps. Table 1 gives a comparison of the main public blockchain platforms that provide the functionality of developing dapps that store the execution rules in smart contracts on the blockchain platform. The blockchain provides the database for storage and retrieval of data generated through dapps. The metrics depicted were retrieved from various sources and correspond to 13$^{th}$ January, 2022.

Table 1 gives the throughput or the number of transactions per second that the concerned blockchain platform is able to achieve and Solana has the maximum throughput with Ethereum being the lowest in terms of performance. The smart contracts can be written in multiple programming languages and blockchain platforms, which provide popular programming languages to write smart contracts like Solana are more attractive for developers. Similarly the transaction fee is an extremely important parameter when developing dapps that will involve a large number of transactions/ users. Stellar has the lowest transaction fee but only the base fee has been indicated in Table 1, while the actual transaction fee would depend on the number of operations a transaction entails. Solana scores in having minimal transaction fee after Stellar.

In consideration of the climate goals outlined in COP26, the consensus protocol used by the blockchain platform should not be energy intensive, so the usage should not contribute to an increase in the number of carbon emissions. Additionally, the hardware requirements for the organization to participate as a validator of transactions must also be researched upon, while taking care of the overall budget of the organization. This scenario would arise if the organization intended to participate in the transaction validation process in the blockchain platform. The market capitalization (or market cap as indicated in Table 1) gives an indication of the investor confidence in the blockchain platform. It is also a prediction of the quality of the technological offering as well as the resiliency of the blockchain platform. A large market cap by default indicates a larger window for the platform to resolve issues in code, for example, and come forward with a sound offering.

*Table 1: Comparison of Blockchain Platforms for Dapp Development*

| Blockchain platform | Throughput (tps)[1] | Smart Contracts Programming language | Consensus algorithm | Market Cap (USD) | Transaction fee (USD) | Public |
|---|---|---|---|---|---|---|
| Algorand | 1000 | TEAL[2] | PoS[3] | 8.8B | 0.0014 | ✓ |
| Cardano | 250 | Haskell | Ouroboros (PoS) | 41.71B | 0.4 | ✓ |
| Ethereum | 13-14 | Solidity | PoW[4] | 389.8B | 6.493 | ✓ |
| Solana | 50,000 | C, Rust, C++ | Tower (PoH[5]) | 46.58B | 0.00025 | ✓ |
| Stellar | 1000 | -NA-[6] | SCP[7] | 6.71B | 0.00000274[8] | ✓ |
| Tezos | 40 | Michelson, SmartPy, LIGO | PoS | 3.7B | 0.00232 | ✓ |

*E. Applications of Dapps in Finance*

The primary benefits of dapps encompass traceability, data immutability, censorship-resistance, service availability, decentralized servers, source code auditability and divisibility of digital currencies being utilized with dapps. Applications of dapps that capitalize on the indicated benefits can be manifold, especially in the finance sector. A few applications have been highlighted in this paper, which can be categorised in the following domains:

*1) Decentralized Finance (DeFi):* Decentralized finance or DeFi comprises of financial products and services developed using the decentralized, distributed ledger infrastructure as in a blockchain [42]. Blockchain absolves the need for intermediaries and financial services are provided through a trustless financial domain in DeFi. Transactions happen in real time and since intermediaries are not required there is cost reduction. This trustless economy is based on the concepts of game theory [43], where validators are given economic incentives to maintain the integrity of the blockchain network. Decentralized borrowing and lending can be accomplished through dapps. NFTs, cryptocurrencies or fungible assets can be used as a collateral. The rise of DeFi has given rise to novel business models [44], like decentralized marketplaces, automated settlement trading, tokenization of Islamic bonds, derivatives trading, liquidity pools and yield farming, among others.

*2) Embedded Finance:* Embedded finance implies the offering of financial services, through non-financial companies by using third-party providers or their own resources, to their clients. Embedded finance provides benefits of time saving within the same physical location to the consumers, who do not have to go to a financial service provider if they need credit but can buy the product they intend to, through the retailer. The embedded financial provider offers a seamless experience to the consumers by integrating the payment mechanism as a part of the service offering itself. Examples of embedded finance are Apple Pay, Google Pay and PayPal. Blockchain provides cost reduction, faster transaction settlement, transparency and an immutable transaction record. All these features of the technology make it a utilitarian technology to implement embedded finance. Every DeFi dapp is an embedded financial offering at its core, where users use cryptocurrencies in exchange for fiat to utilize the decentralized financial offering. Embedded finance coupled with blockchain-based dapps can offer benefits in multiple areas like metaverse, digital commerce, insurance and even gaming [45].

*3) Web 3.0:* Web 1.0 offered users the ability to read data, whereas Web 2.0 enabled users to not just read but also write data, contributing to the available content for consumption. Web 3.0 extends Web 2.0 and facilitates the users to read, write and execute implying participation by the general population in contributing to the backend of software applications [46]. This contribution enables decision-making to be transferred to the average user and encompasses incentives through monetary rewards for the service provided by the user. This forms the foundation of tokenomics [47], which is powered by blockchain. NFTs would pave the way for assigning rights and privileges in the digital world, equivalent to ownership rights in the physical world. Metaverse could be powered by a new cryptocurrency. The integration of blockchain, AR (Augmented Reality), VR (Virtual Reality), MR (Mixed Reality) and NFTs [48] will change the way businesses are conducted and cause disruption in the financial sector. However, the concept is still in its infancy.

*4) Insurance:* The insurance industry has been slower to adapt a tech model, in contrast to their companies in retail banking and capital markets, but in hindsight it can result in a massive advantage on account of the wider acceptance of the digital landscape and maturity in underlying technologies. Insurance is now ripe for disruption as reflected in the fast growth of insurtech deals [49]. Blockchain is a pivotal technology that can help to automate tasks in the insurance industry [50]. An example would be automating claim settlement through smart contracts that get triggered on the occurrence on an event for which an insurance cover exists. Dapps that synchronize the activities associated with claim settlement and harmonise the actions of various intermediaries through a common, decentralized ledger for data sharing can revolutionize the insurance industry. Other applications also exist, whereby dapps can provide authentic and non-repudiated data for AI (artificial intelligence) algorithms to offer customized insurance products for example to the users. Blockchain can help to prevent fraud [51] and through the employment of machine learning predict the occurrence of events for each insured individual, based on metrics derived from their life.

---

[1] transactions per second
[2] Transaction Execution Application Language
[3] Proof of Stake
[4] Proof of Work
[5] Proof of History
[6] Does not have a smart contract programming language
[7] Stellar Consensus Protocol
[8] Base fee

*5) ESG*: Investors observe a correlation between the financial performance of a company and their navigation of environmental challenges/ opportunities as per a study conducted by Research in Finance [52]. Climate change, carbon emissions and pollution rank as the major environmental concerns amongst investors. Independence of board, executive remuneration and diversity in the board composition were foremost in the governance criterion when investors evaluated a firm. Rights of workers, equality and diversity were important factors for investors when looking at the social considerations [53]. When we focus on the environmental facet of ESG then an authentic impact assessment and measurement becomes extremely important. Blockchain through dapps developed to accommodate the ESG criteria come across as invaluable tools to provide a non-repudiated account of the environmental progress achieved to thwart climate change or reduce pollution. The European Commission for example intends to use blockchain to fight climate change [54]. Blockchain through the use of smart contract, oracles and dapps can interact with real-world data to cope with the impact of climate change [55]. In accomplishing transparency in governance, blockchain through smart contracts and dapps can facilitate honest voting on decisions, where all concerned stakeholders can participate with their votes permanently recorded on the blockchain [56]. Dapps that support financial inclusion and provide tokens to record the impact of a social initiative leverage on blockchain for a quantitative measurement of the benefits that are being achieved socially [57].

## IV. DCARBONX: DECENTRALIZED CARBON MARKET APPLICATION SOFTWARE

DcarbonX, Nash fintechX's proprietary software, brings to market NFT-based carbon credit tracking and a secondary marketplace. DCarbonX facilitates secure, real time logging and tracking of carbon credits, while functioning as a marketplace on its integrated platform. The DCarbonX platform can be accessed by SMEs as a common marketplace or can be built on top of existing internal systems in large enterprises and government institutions. The functional architecture of DCarbonX is described in Subsection IV-B.

The company intends to launch the product in the market in Q3 of 2022 as well as to homologate the solution with climate bodies in the main target regions e.g. European Climate Change program, CAMENA and Task Force on Voluntary Carbon Markets (TSVCM). The target market is VCMs (Voluntary Carbon Markets) and NDC (Nationally Determined Contributions) projects, funded by conventional financial institutions and Islamic financial institutions. While, no official estimation of market exists, Nash fintechX's internal analysis shows that a target $CO_2$-equivalent reduction of 23 $GtCO_2$-equivalent (including 2.5 $GtCO_2$-equivalent for VCM project) by 2030 translates to projects worth $66B. Assuming high adoption of blockchain solution in VCM projects and low-medium adoption in NDC projects, this translates to $30-40B worth of transactions annually (2030-2050) on logging impact and carbon credit sales on the DCarbonX platform. The company intends to work closely with banks, financial institutions and Islamic financial institutions for project definition, tracking as well capacity building of internal teams on ESG impact topics. There is a major challenge to get the stakeholders aligned on this novel approach, including climate bodies, clients executing the project and project auditors. The main revenue will be generated from the transaction fees from NFT issuance from the platform and project consultancy.

*A. Motivation*

COP26 was a landmark event in the fight against climate change as all 196 countries at UN signed the international treaty on climate change mitigation, commonly known as Paris accord. The primary aim is to keep the rise in mean global temperature to below 2°C and preferably to 1.5°C. The Paris accord also foresaw countries to build national plans known as Nationally Determined Contributions (NDCs) on emission reduction targets every 5 years. The COP26 held in Glasgow in October-November 2021 saw 194 countries presenting their first NDCs while 143 of them submitted new or updated plans in time for COP26. These new, updated plans would lead to total GHG (greenhouse gas) emissions being about 9% lower in 2030 than they were in 2010. The UN estimates that the current commitments align with a temperature increase of 2.7°C by the end of the century. This will mean additional commitments on emission reduction on part of member countries to reach Paris accord target of average temperature increase of 2°C or preferred target of 1.5°C. COP26 was also the most significant COP for companies and the finance sector, reflecting the growing expectations of both to play key roles in enabling decarbonization in accordance with the Paris goals.

One of the major points of discussion at COP26 related to Article 6, which sets forth a framework for the creation of a voluntary international carbon credit trading market. The article 6.2 foresaw that countries will voluntarily engage in "cooperative approaches" involving internationally transferred mitigation outcomes ("ITMOs", emissions credits transferrable across globe) towards their NDCs. The Article 6.2 also provides that, in so doing, countries must apply robust accounting to ensure avoidance of double counting of emission credits [58].

The goals of article 6.2 are operationalized in article 6.4, which defines a "mechanism" supervised by a body designated by the parties to the Paris Agreement to:

1) Promote the mitigation of GHG emissions
2) Incentivize and facilitate the mitigation of GHG emissions by public and private entities

Although Article 6 lays the foundation for the creation of a global carbon credit market, it left key implementation details to be determined later. There were differences on structural topics including mechanism to avoid double counting and trading of older carbon credits. After five years of deliberations, there was fortunately an agreement at COP26 on the 'rulebook' for this carbon credit market for the states and private entities to generate and trade carbon offset credits. While the agreement is a leap forward for transparency on country level NDCs achievements and carbon credit transfer, the big concern remains on lack of mechanisms governing private projects outside a country's NDC or not adjusted for in the country's carbon budget. These emission reductions certificates can still be sold to corporate emitters or other non-state actors, who could use them to claim that they are reaching "climate neutrality" without a robust tracking and logging mechanism. Currently only few voluntary carbon markets are established for entities to buy and sell carbon offset credits at their discretion (e.g., to fulfill a voluntary commitment to reduce emissions). Lack of standardization in these voluntary markets has raised numerous concerns, including with respect to the quality and validity of offsets and the associated credits [58].

There has been a flurry of announcements on voluntary contribution on carbon emission reduction in the run-up to COP26 by private entities, both in high carbon emitting sectors as well as financial sectors. The climate targets will become a key performance metric and competitive advantage for players in their respective business. Thus a robust, secure, transparent carbon tracking and trading system is a big business opportunity in the run-up to the first climate change milestone of 2030. DCarbonX aims to fulfil this exploding market need with its blockchain-based emission tracking and trading platform as a market pioneer [59].

*B. Functional Architecture*

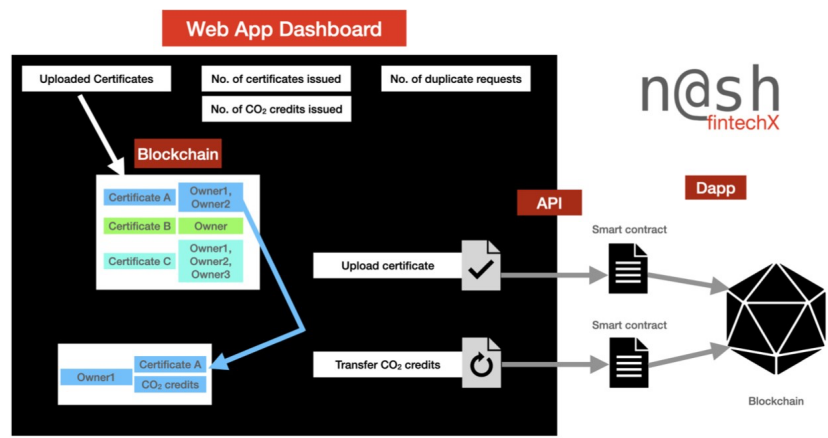

*Figure 5: Functional Architecture of DCarbonX.*

DCarbonX is a blockchain-based carbon market. It is a decentralized application software that utilizes blockchain to store an immutable record of issued and traded carbon credits. It provides a primary and secondary market for trading of carbon credits, stored as NFTs on the blockchain platform. These carbon credit are created as NFTs following a similar methodology as was employed for Figure 3.

Figure 5 represents a simplified functional architecture of DCarbonX. The user through the organization that uses DCarbonX accesses the NFTs of carbon credit certificates through a dapp. Smart contracts help to create NFTs of uploaded carbon credit certificates and also facilitate trading of NFTs. The organization and the users can see analytics related to the issued carbon credit certificates, where each carbon credit certificate would list down the owners, indicating the original owner and then all the subsequent buyers in the trading market. Additionally, the dashboard developed through a web app working in integration with the dapp will depict the total number of issued carbon credit certificates, total number of uploaded carbon credit certificates or NFTs and the number of duplicate requests whereby someone tries to upload a carbon credit certificate that already exists on the blockchain.

*C. Minimum Viable Product*

Figure 6 depicts screenshots from the minimum viable product (MVP) of DCarbonX, which has been developed on Solana [60] blockchain platform. The screenshot on the left shows a carbon offset NFT on Solana blockchain. The certificate is stored on Arweave [61], while the hash is stored on Solana. The hash and the related transaction can be seen by clicking the keyword 'Solana' while to traverse the uploaded certificate, 'Arweave' needs to be clicked. The MVP uses Phantom wallet [62], which can be installed on Chrome browser. The Phantom extension helps users to create a blockchain-based wallet, buy and store the cryptocurrency of Solana and helps in transaction payments. The screenshot also shows the profile, account balance of the concerned user and permitted functions for that profile, which are to create an NFT, sell an NFT, buy cryptocurrency and disconnect from the Phantom wallet. DCarbonX can be developed on any blockchain platform providing smart contracts but our analysis revealed that Solana would be an optimum choice considering our need that the blockchain platform should not be energy intensive like Ethereum, have a high throughput and should have minimal transaction fee. Our comparison of blockchain platforms in Table 1 on which dapps can be developed supports our conclusion in favor of Solana. The screenshot on the right shows the listing of NFT in the primary market in the Solana blockchain, which can be reached by clicking the keyword 'Solana' in the screenshot on the left. This provides an immutable record of the carbon offset as an NFT on the blockchain. The DCarbonX dapp provides a drag-and-drop GUI (graphical user interface) so the users visualize the carbon market as being similar to the existing web apps in simplicity of design, while the backend complexity is hidden from them.

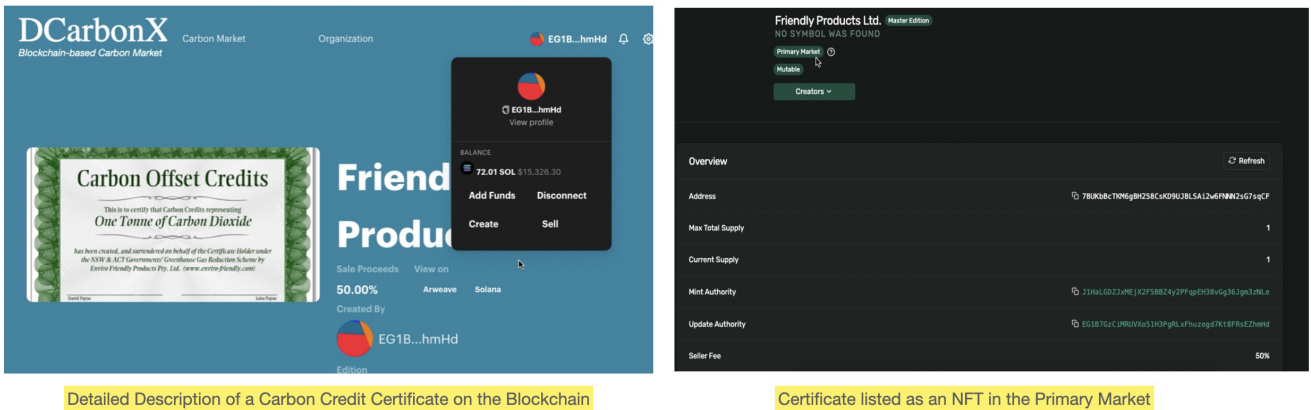

Figure 6: Screenshots from the MVP of DCarbonX.

*D. Salient Features*

The salient features of DCarbonX together with the advantages it provides can be summarized as follows:

1) DCarbonX is under the control of the issuing organisation and only registered entities can participate in the carbon market.
2) The issuing organisation has the sole power to create NFTs whereas all other users can buy and sell them. This can be extended to accommodate other issuers, if needed.
3) A carbon credit NFT issued by the organisation is listed in the primary market.
4) Any purchase and resale of the issued carbon credit NFT lists it in the secondary market.
5) Interface to view all issued carbon credit certificates through the web app dashboard.
6) Analytics indicating total number of issued carbon credits through NFTs.
7) The NFT of a carbon credit certificate has its entire list of owners listed on the dashboard.
8) The issuing organization gains a certain percentage of the profit each time the issued carbon credit NFT is sold in the carbon market.
9) At any instant of time all the transactions are available on the public blockchain for validation.
10) The identity of the registered users is known only to the issuing organization to ensure privacy, whereas transaction records are public for transparency.
11) An immutable record of issued carbon credit NFTs will prevent duplication of records, thus preventing greenwashing.
12) All claims made by an organization related to its positive contribution to achieving carbon neutrality will be backed by authentic and non-repudiated records on the blockchain.

13) DCarbonX has a GUI, which hides the complexity of blockchain from the users.
14) The overall infrastructure will provide the requisite accountability, transparency and trading of carbon credits to strengthen the efforts of achieving the climate goals outlined in COP26, adhering to a structured and quantifiable approach.
15) Potential deployment of AI on the blockchain data to offer recommendations to traders on which carbon credit NFT to buy, indicate the best price for sale and predict demand.

*E. Utility in Sustainable Finance*

One of the biggest challenge in achieving the 1.5°C Paris accord goal remains the funding for climate projects contributing to NDCs and VCMs. A commitment of $100B per year by donors to support low and middle-income countries in tackling climate change was among the key goals of COP26. However, there was no explicit agreement at COP26 on the mechanism to fully deliver this commitment upto 2025. There has been a push to mobilize private sector financing to bridge this gap and more than 450 firms totaling $130T in asset base have formed the Glasgow Financial Alliance for Net Zero (GFANZ), a forum for financial institutions to accelerate the transition to a net-zero global economy. The GFANZ will report progress on financed projects but will also define net zero pathways for major sectors, thus aligning efforts from corporates and financial institutions. Large private sector banks have joined the fray with Goldman Sachs committing $750B to sustainable finance over the next decade. This followed an earlier announcement from JPMorgan Chase to facilitate $200B in sustainable finance in 2020. There still remains a financing gap which has to be bridged by innovative, sustainable financing asset classes. Green bond has emerged in recent years as a major sustainable financial tool with the issuance touching $500B in 2021. However, continued growth of green bonds remains incumbent to solving two major challenges associated with the ESG sector, namely greenwashing and higher issuance costs linked to impact assessment and reporting.

The Islamic principles of moderation in consumption, and avoidance of the wasteful use of natural resources makes faith an integral force multiplier in the fight against climate change. This is especially important as 57 member countries of OIC are home to almost 24% of world's population with many of the member countries likely to be heavily impacted by the effect of climate change. The Islamic finance products with it's unique asset-backed, ethical and risk sharing principles lends itself naturally to principles of sustainable finance. There has been a greater focus on **Green Sukuk** as a major asset class to meet climate financing needs. The market for green sukuk has grown rapidly with issuance of $3.5B in 2019, driven partly by increased liquidity in investor base in MENA and Asia. There has been a diversification in investor base with a push for ESG products in EU and other developed regions. However, like its counterpart in conventional finance, green bond, green sukuk also involves rigorous project monitoring and impact logging making it difficult to catch up the market appetite. Thus, creating a supply gap. DCarbonX aims to solve the problem through early involvement in project selection as indicated in Figure 7 and supporting the project teams throughout the project lifecycle.

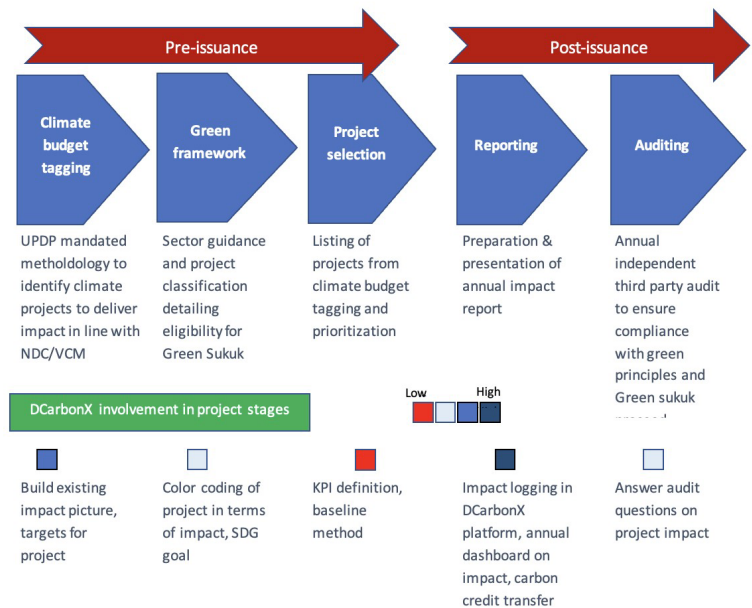

*Figure 7: DCarbonX and IsDB Green Sukuk Issuance Cycle.*

## V. CHALLENGES IN DEVELOPMENT, DEPLOYMENT AND USAGE

Blockchain technology, despite being revolutionary for the finance sector, has still not seen mass adoption. The advent of NFTs witnessed an increase in the user base and brought the technology to the realm of the common man for an asset-backed productive usage, instead of being an investment vehicle for cryptocurrencies. However, there still exist challenges that need a coordinated effort from all stakeholders to deploy the technology to solve existing challenges, for which an alternative solution does not exist like for climate change. Some of the challenges related to dapps, demarcated into different categories, have been enumerated below:

*A. Development*

- **Limited blockchain experts.** Blockchain technology is still in the stage of development with most projects still in the pilot phase. On account of this people pursuing advanced studies/ skills in the technology are comparatively few in number. This has led to scarcity of requisite talent, leading to a limited number of blockchain experts.
- **Absence of smart contract auditors.** Smart contracts developed on different platforms form the foundation for dapps. Any transaction emanating through dapps is permanent with the resulting data leaving an immutable footprint in the blockchain platform. Hence the code for the smart contract should be critically analysed for any potential issues and even security flaws, which can lead to drainage of funds from a user's wallet or other similar cybercrimes. This analysis can be accomplished by smart contract auditors who are well aware of the programming language and the various issues associated with the use of blockchain. This knowledge in multiple areas has created a need for experts, who are relatively fewer in number.
- **Scarce developer community.** Blockchain platforms employ different programming languages to build smart contracts, including many new languages native to a particular blockchain platform like Solidity in case of Ethereum. This has led to the need for developers proficient in these languages. Lack of mass adoption has prevented many developers, from venturing into the area. Thus, the developer community in blockchain is small. Additionally, the complexity associated with the development of a dapp, which requires development skills in not just the requisite blockchain platform but also an understanding of the technology to ensure robust smart contracts that back dapps, has attracted relatively fewer developers.
- **High initial capital requirement.** In consideration of scarcity of blockchain experts, smart contract code auditors and blockchain developers, the development community is in a lot of demand and expense incurred to deploy a blockchain-based solution is very high. Moreover, there needs to exist cooperation between developers concerned with different facets pertaining to dapp development like the frontend, smart contract coding, testing, and integration with a centralized database/ oracles if needed. This lends complexity to the process requiring comparatively more human resources to bring to realization the target dapp as compared to a web or a mobile app. Thus, any venture that undertakes the task of developing a dapp requires a very high initial capital to integrate and employ the services of requisite experts and developers.
- **Lack of supporting tools.** The smart contract technology is still new and has not attained maturity. Consequently tools to support the development of smart contracts and dapps that simplify the process and help to accomplish the tasks efficiently are still work in progress or non-existent depending upon the blockchain platform.
- **Inherent bottlenecks in the blockchain platform.** Blockchain platforms might deal with different obstacles based on the consensus mechanism employed by them, among other factors. Scalability and throughput are dominating issues in Ethereum on account of proof of work consensus. Additionally, it involves considerable expenditure of energy, like in Bitcoin and many other privacy-preserving blockchain platforms like Monero, Zcash and Dash, which makes it unsuitable to be used for ESG purposes. Throughput is a major issue in other blockchain platforms as well, where it is seen that the transactions per second, as given in Table 1, in most blockchain platforms is not sufficient to support a large number of users in parallel. Credit card companies like Mastercard can settle 5000 transactions per second [63].

*B. Deployment*

- **Integration of KYC.** KYC or *Know Your Customer* is a mandatory requirement in financial services to help validate the identity of the user and authenticate the business relationship. KYC solutions are provided by third-parties or they can be developed in-house, which results in an increase in the total costs. The dapp needs to work seamlessly with the KYC services already provided by an organisation. Alternatively, if KYC process is not existent in an organisation, then they should be integrated during dapp development to ensure that the mandatory requirements are catered to for dapp usage.
- **Integration with existing legacy systems.** In majority of solutions where the dapp caters to deployment by organisations and is not meant for direct retail use, there needs to be an integration with the existing infrastructure of the organisation. This integration can be with the organisation's existing client registration processes, access and retrieval of data from centralized databases and updating of existing database in the organisation to populate it with blockchain data for example. This integration requires expertise in existing web technologies together with blockchain.
- **Inadequate regulatory support.** Regulatory support is crucial for the sustainable progress of innovations. Regulations have an impact on innovation [64] and their existence is necessary to protect the users from malpractices and uncertainties associated with a novel technological implementation. Blockchain technology, cryptocurrencies and hence dapps do not have the requisite regulatory support in most countries that is essential for their secure deployment and usage [65].
- **Non-uniformity in wallet providers.** Regional differences exist as to the permissibility of utilizing a blockchain-based wallet for a dapp, with some countries permitting a certain wallet provider, whereas other countries prohibiting it. This poses an additional constraint during development as the dapp needs to be adapted according to the geographical region of it's offering. Different wallet providers have a different fee structure and even a different methodology for onboarding of clients.

*C. Usage*

- **Security issues and errors in code.** Vulnerabilities in smart contract code like coding errors can be exploited to make the usage of the concerned dapp insecure. Once the smart contract is deployed on the blockchain, the code cannot be updated so any bugs that exist would lead to undesirable effects. The vulnerability in code led to the loss of $50M in the DAO attack on Ethereum [66].
- **Requirement of cryptocurrencies.** All blockchain platforms function by charging a certain fee for the utility they provide. This fee must be paid in the native cryptocurrency of the blockchain platform. So any user that utilizes a dapp built on a public blockchain needs to have the native cryptocurrency in their wallet that would make transaction payment possible for the blockchain platform. There can be many dapp architectures where organisations can go for a private or a permissioned blockchain and resort to using fake tokens in exchange for fiat currency, doing away with the need of cryptocurrencies. However, in general whenever a public blockchain platform is used, which is an optimum choice whenever immutability and transparency is at the core of an offering like the ESG sector, the user would need cryptocurrencies. Another alternative can be developed, whereby the organisation pays in the cryptocurrency on behalf of the clients. However, the current level of advancement that exists in the technology as of now, envisages that a user requires cryptocurrencies to use a dapp on a public blockchain. This is a challenge as fiat currency is universally accepted and so their procurement and management tools are available but cryptocurrencies are a recent embarking and do not have the requisite support as of now.
- **Lack of trust in the technology.** Blockchain is still considered to be a novel technology. The initial domination of cryptocurrencies as an investment vehicle, which witnessed massive fluctuations in prices and led to multiple scams through ICOs has created suspicion in the average user and organisations [67]. The usage of cryptocurrencies in cybercrime and the dark web had added to the lack of trust in the ethical nature of the technology [68]. Moreover, the scalability and throughput issues that plague the technology [69] have garnered a public opinion that associates lack of efficacy with the technology, rendering it as a technology that is still in research and development phase.
- **Lack of technical literacy.** Users are not educated on the various complexities associated with the use of blockchain and dapps. There is lack of awareness of the technology itself, with the existing consumers limited to entities with the necessary technical knowledge required to use it. However, mass adoption would occur realizing the full benefits that the technology offers, once the level of technical literacy that is required and the ease of usage is similar to how users interact with existing apps on the web or mobile.
- **Management of keys.** The public and private keys associated with the blockchain wallet need to be kept secure and different management strategies exist like a web wallet, hot wallet or a cold wallet, among others. This imposes an additional burden on the users in terms of technical awareness.

## VI. CONCLUSION

Decentralized applications developed utilizing blockchain technology capitalize on and provide all the benefits that accrue with blockchain usage. Dapps pave the way for novel business models to be developed that can impact the economy positively, considering the focus on a resilient digital infrastructure, instigated by the problems during the coronavirus pandemic. In this paper we embark on a case study of a decentralized application software, DCarbonX, which provides the necessary support to mitigate the harmful effects of climate change and provides a quantifiable approach to achieving carbon neutrality by 2050, as per the outlined goals in COP26. We give the related work on dapps and highlight that a parallel to the present work has not been delved upon so far to the best of our knowledge.

The paper gives a comparison of blockchain platforms that can be utilized to develop dapps and highlights the applications of dapps in pertinent areas of DeFi, embedded finance, insurance, Web 3.0 and ESG.

The main contribution of the paper is the case study of DCarbonX, which depicts the practical implementation of a decentralized application in the very relevant area of climate change. Climate change remains the biggest challenge confronting human race. Climate projects pose the possibility of greenwashing and duplication of impact. Even for NDC projects there is an elaborate bureaucratic process on logging and auditing the impact. The current practices thus add to the cost of financing of climate projects, while not contributing to the cause of transparency and fast deployment of projects. The paper highlights how through DCarbonX, climate projects can be made traceable, their impact can be recorded and how a transparent carbon market can be developed using blockchain.

The paper gives the functional architecture of DCarbonX and defines the salient features of the MVP of DCarbonX. The paper also sheds light on the critical role sustainable finance will play in execution of climate projects and specifically the possibility to expand green sukuk for meeting the financing gaps. The section also details integration of DCarbonX in the green sukuk issuance cycle to ensure a good product design and robust impact tracking. The paper is a pioneer in describing the challenges involved in development, actual deployment and usage of decentralized applications.

## ACKNOWLEDGMENT

The authors would like to thank Dr. Rushdi Siddiqui for his support and invaluable guidance in creating the promotional video for DCarbonX, which can be accessed using the YouTube link, https://youtu.be/bBDFC42eq1s.

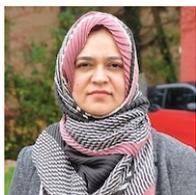

**Nida Khan** is the Founder and CEO of Nash fintechX, Luxembourg. The company provides software solutions, specializing in blockchain and AI. She was ranked as the 11$^{th}$ most influential woman in Islamic Business and Finance in 2021. She is also working as a consultant on a project on Arab Vision by United Nations Economic and Social Commission for Western Asia. She holds a PhD in computer science from the University of Luxembourg, Europe and a Master Diploma in Islamic Finance from AIMS, UK. She was accorded the prestigious FNR (Luxembourg National Research Fund) grant, given to innovative industrial projects, for her doctoral research on blockchain in finance. She has worked on projects in social finance and decentralized finance, including tokenization. She is a pioneer in developing a novel consensus mechanism for scalable blockchains and ensuring GDPR-compliant privacy preservation in blockchains. She has worked with Tabrez in multiple projects, including pioneering contributions related to giving the economic impact of blockchain-based micropayments and proposing a mathematical framework for blockchain governance. Her past achievements include developing two novel iOS and Android apps, where the latter had 10K+ installs. Her work appears in diverse publications and she speaks regularly on emerging technologies. Her current research interests include artificial intelligence and quantum computing.

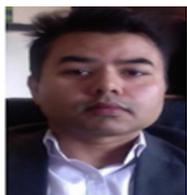

**Tabrez Ahmad** holds a Bachelor's degree in Chemical Engineering from IIT Kanpur, India and a Master's degree in Business Administration from IIM Calcutta, India. He is currently working in ArcelorMittal, Europe as the Product Leader for flat steel products, where he is responsible for the marketing function in Europe. His prior work experience involves contributing to Internal Consultancy and Corporate CTO functions at ArcelorMittal group. Besides his extensive experience in the Metal Sector, he has also worked in an Oil and Gas major in Sales and Marketing roles. Tabrez has worked with Nida in developing a mathematical model for blockchain governance based on Nash equilibrium. The model can be used to predict with accuracy the future growth and majority support in a blockchain fork, based on the voting results. He has contributed to the analysis of Stellar blockchain data from a micropayments' perspective. He, with Nida, is a pioneer in giving the economic impact of blockchain-based micropayments. He recently developed a calculator for deriving the carbon footprint of an entity, supporting it with the requisite background information to lend transparency and provide standardization to the process. His current research interests include artificial intelligence and blockchain.